\documentclass[aps,prm,groupedaddress,amsmath,amssymb,longbibliography,twocolumn]{revtex4-2}

\usepackage{graphicx}
\usepackage{dcolumn}
\usepackage{color,soul}
\usepackage{lipsum}
\usepackage[colorlinks=true, citecolor=blue, urlcolor=blue]{hyperref}
\usepackage{xr}
\usepackage{lineno}
\usepackage{longtable}

\usepackage{color,ulem}

\begin{document}

\title{All ``roads'' lead to rocksalt structure}

\author{Matthew Jankousky}                                                                                                     
\affiliation{Colorado School of Mines, Golden, CO 80401, USA}

\author{Helen Chen}
\affiliation{Harvey Mudd College, Claremont, CA 91711, USA}

\author{Andrew Novick}                                                                                                     
\affiliation{Colorado School of Mines, Golden, CO 80401, USA}

\author{Vladan Stevanovi\'{c}}
\email{vstevano@mines.edu}                                                                     
\affiliation{Colorado School of Mines, Golden, CO 80401, USA}

\date{\today}

\begin{abstract}
Rocksalt is the most common crystal structure among binary compounds. Moreover, no long-lived, metastable polymorphs are observed in compounds with the rocksalt ground state. We investigate the absence of polymorphism and the dominance of the rocksalt structure via first-principles random structure sampling and transformation kinetics modeling of three rocksalt compounds: MgO, TaC, and PbTe. Random structure sampling reveals that for all three systems the rocksalt local minimum on the potential energy surface is far larger than any other, making it statistically the most significant structure. The kinetics modeling shows that virtually all other relevant structures, including most of the 457 known A$_1$B$_1$ prototypes, transform rapidly to rocksalt, making it the only option at ambient conditions. These results explain the absence of polymorphism in binary rocksalts, answer why rocksalt is the structure of choice for high-entropy ceramics and disordered ternary nitrides, and help understand/predict metastable states in crystalline solids more generally.   
\end{abstract}

\maketitle

\begin{figure*}[t!]
\includegraphics[width=\linewidth]{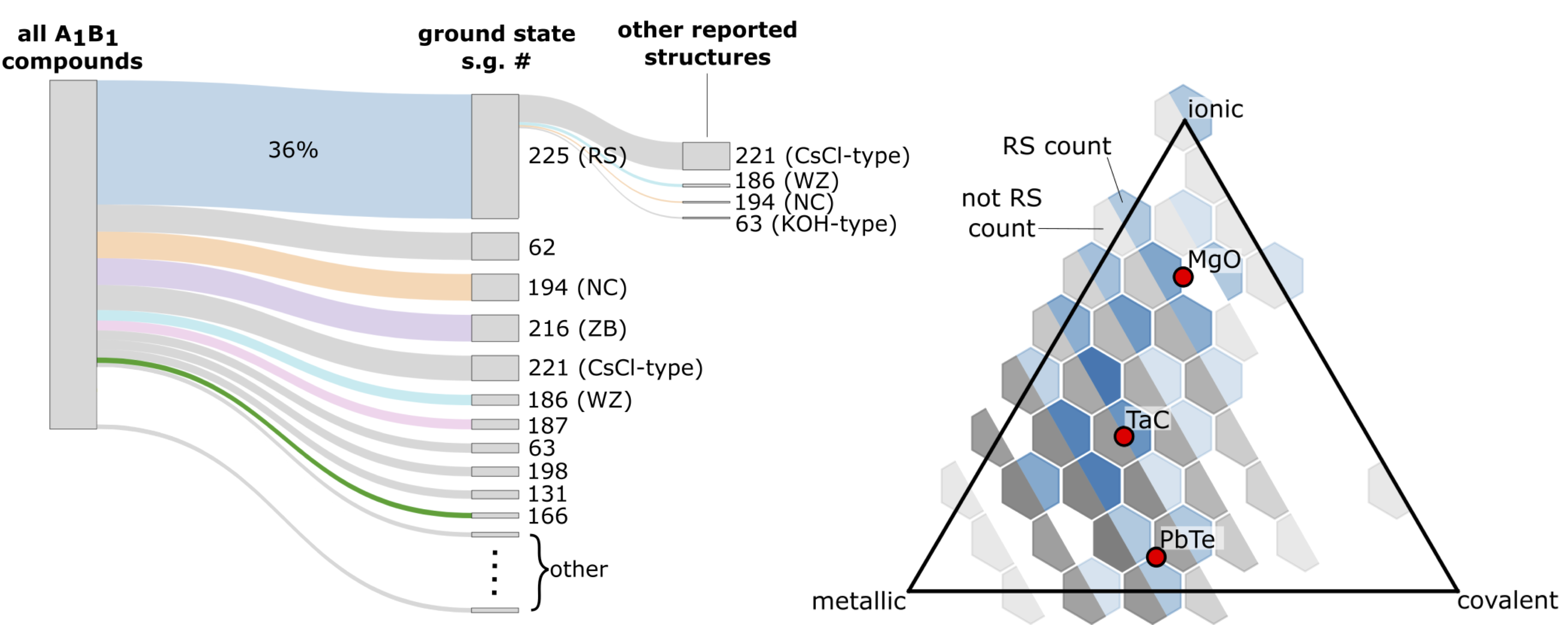}
\caption{\label{fig:abundance_fig} 
(left panel) A sankey diagram showing the distribution of ground state structures (their space group numbers) among known A$_1$B$_1$ compounds. Other reported short-lived polymorphs of compounds with the rocksalt ground state are also shown. (right panel) The Van Arkel-Ketelaar bonding triangle populated with known A$_1$B$_1$ compounds. Darker hexes imply more compounds. Those with the rocksalt ground-state are highlighted in blue with the three compounds studied in this work depicted by red circles.
}
\end{figure*}

The rocksalt crystal structure dominates both binary compounds and entropy stabilized multi-component systems  \cite{NitrideMap,ZnZrN2,RocksaltNitrideSemiconductor,CurtaroloNatureCeramics,CurtaroloNatureOxides, CurtaroloNatureCarbides,NavrotskyHEOThermo}. Within the A$_1$B$_1$ chemical space, which spans tetrahedrally bonded semiconductors (GaAs, SiC, GaN, ZnO,\dots), photovoltaic materials (CdTe), piezoelectrics (ZnO), thermoelectrics (PbTe), superconductors (NbN, FeSe) and other important functional materials, rocksalt is the most common crystal structure. There are 348 compounds with the rocksalt structure reported in the Inorganic Crystal Structure Database (ICSD) \cite{icsd} out of 704 total A$_1$B$_1$. Moreover, rocksalt is the ground state structure in the Materials Project database (MP) \cite{materialsproject} for 238 of total 668 A$_1$B$_1$ compounds found in MP, as illustrated in Fig.~\ref{fig:abundance_fig}. In contrast, the next most frequent A$_1$B$_1$ structure, zincblende, is the ground state for only 51 compounds. 
Rocksalt is also the most frequent structure among binaries as a whole. The next most common binary structures are fluorite (also s.g.~\#225) with 75 A$_1$B$_2$ compositions and the Uranium silicide-type structure  (s.g.~\#221) with 66 A$_1$B$_3$ compositions. Regarding the entropy stabilized systems, disordered ternary nitrides often crystalize in the rocksalt structure \cite{NitrideMap,ZnZrN2,RocksaltNitrideSemiconductor}. High-entropy ceramics (oxides, nitrides, and carbides) with a 1:1 metal to nonmetal ratio always form in the rocksalt structure even when many parent compounds in an equimolar mixture are not rocksalts themselves \cite{CurtaroloNatureCeramics,CurtaroloNatureOxides,CurtaroloNatureCarbides,NavrotskyHEOThermo}.  

\begin{figure*}[ht!]
\includegraphics[width=\linewidth]{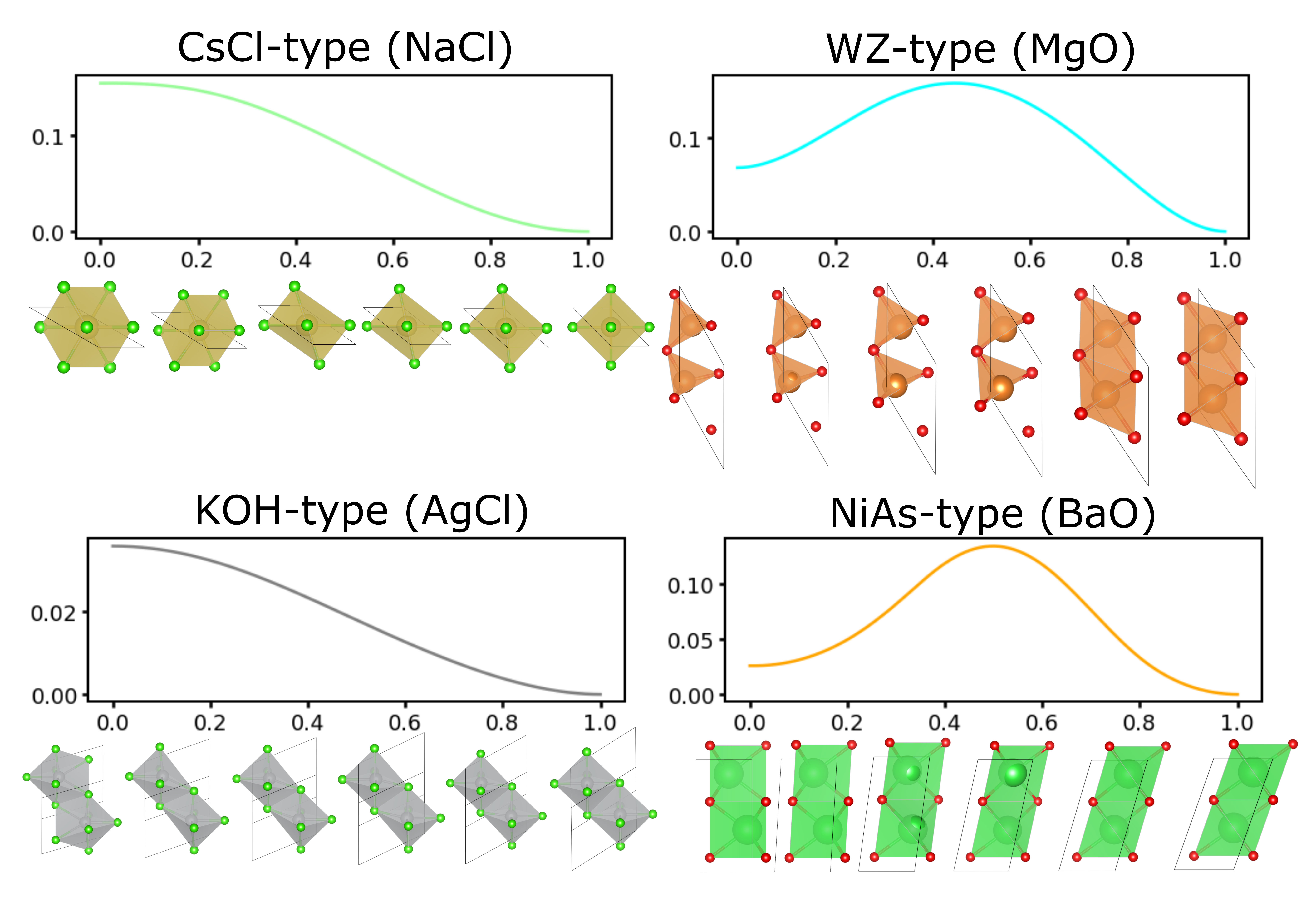}
\caption{\label{experimental_paths} Computed pathways and barriers for experimentally reported polymorphs of rocksalt compounds: CsCl-type in NaCl, NiAs-type in BaO, KOH-type in AgCl, and wurtzite-type in MgO.}
\end{figure*}

While it is commonly associated with ionic compounds, the rocksalt ground state appears across the entire Van Arkel-Ketelaar bonding triangle \cite{bondtriangle} shown in Fig.~\ref{fig:abundance_fig}, displaying that this structure is favorable in a variety of chemical environments. Arguments based on maximizing the packing factor for a given combination of ionic radii \cite{packingfactor} or Phillips' descriptor that chemically delineates which A$_1$B$_1$ systems will adopt a rocksalt structure instead of a wurtzite structure \cite{ionicity} might help explain why rocksalt is the ground state structure. However, neither of these concepts is sufficient to predict or explain whether rocksalt compounds will have metastable polymorphs or not. 
%
%
%
%
%
For all the abundance and diversity of compounds with the rocksalt ground state, there are virtually no experimentally reported metastable polymorphs of A$_1$B$_1$ rocksalts in MP \cite{materialsproject}, ICSD \cite{icsd} or available literature. 
The only reported cases are high pressure phases of the CsCl-type \cite{Abrahams1965}, NiAs-type \cite{Weir1986}, and KOH-type \cite{Hull1999}, and wurtzite in extremely thin films \cite{MgOWZthinfilm,NamKiMi2012} (see all compounds in Supplementary Table \ref{RS_polymorphs}). None of these 4 structures types represent long-lived, metastable states, as they all transform quickly to the rocksalt structure if the stabilizing pressure or strain is released (Supplementary Figure \ref{experimental_paths}). 

To elucidate the physical origins behind these observations, we investigate three specific systems with rocksalt ground state and distinct bonding characters: ionic MgO, largely metallic TaC, and covalent PbTe. Our analysis derives from the first-principles characterization of: 
(a) the sizes of the basins of attraction of different potential energy local minima (their ``widths''), and (b) the transformation pathways and associated kinetic barriers for polymorphic transformations (``depths'' of local minima). The ``widths'' of local minima correlate to experimental realizability, while the ``depths'' are connected to the lifetimes of polymorphs, and, as we will show, it is the interplay of these two features that explains the absence of polymorphism in rocksalt binary compounds and also its dominance in entropy-stabilized systems.

\begin{figure*}[ht]
\includegraphics[width=\linewidth]{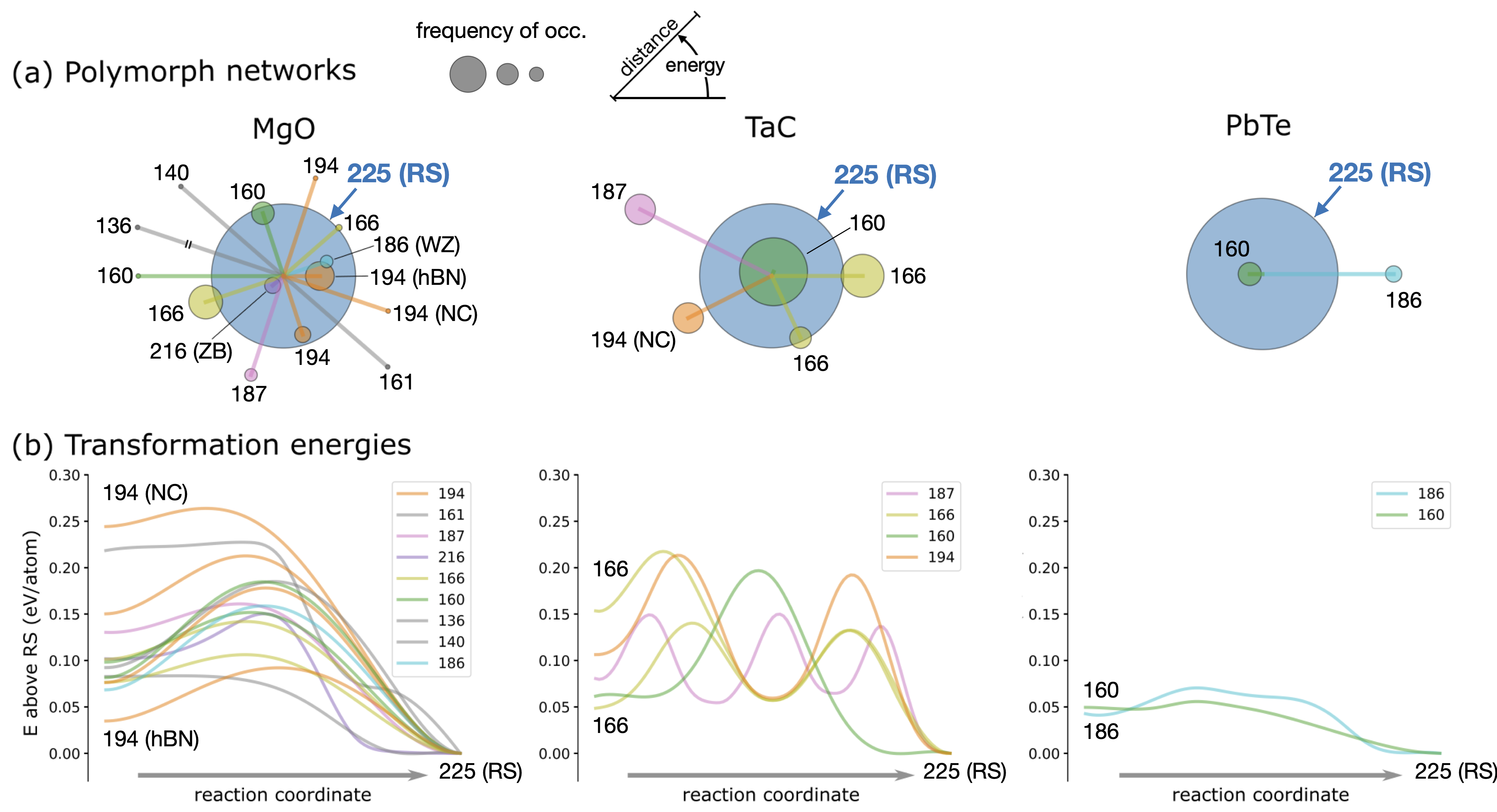}
\caption{\label{fig:random_sample} 
(a) Networks representing the high-symmetry polymorphs identified by random sampling in MgO, TaC, and PbTe. The rocksalt structure is the central node of each network. All other structures are labeled according to their space group number.  The size of each node represents the frequency of occurence of a structure in the random sample. Note that a fixed size is used to represent the width of the rocksalt local minimum, and other points are scaled relative to this size.
Structures are arranged so that the energy increases with the polar angle and the distance of each node from the rocksalt center is proportional to the distance travelled by the atoms during phase transformation of that structure to rocksalt. (b) Calculated (SSNEB) energy profiles along the transformation from the high-symmetry structures identified in the random sampling to the rocksalt structure.}
\end{figure*}

The ``widths'' of local minima are evaluated by the frequency of occurrence of a structure in the first-principles random structure sampling \cite{randomsuperlattice}, which involves generating a large number of initial structures with random unit cell vectors and random atomic positions. Using density functional theory (DFT) \cite{DFT}, those random structures are relaxed to the closest local minimum on the potential energy surface. The frequency of occurrence of a given structure after relaxations then measures the size of its attraction basin, that is, the total volume of configuration space occupied by that local minimum. Statistically, this frequency of occurrence represents the probability to fall into a given local minimum \cite{Sipoly}. It has been shown that the experimentally realized metastable polymorphs are exclusively those with high frequencies of occurrence in the first-principles random structure sampling \cite{randomsuperlattice,SnN,Sipoly,ZnZrN2,polysamplercarbides}. 

The ``depths'' of local minima are correspondingly measured by predicting mechanisms and associated energy barriers for diffusionless transformations between various structures. Mechanisms are identified by mapping crystal structures atom-to-atom in a way that minimizes the distance traveled by atoms along the pathway \cite{pmpaths}. Energy barriers are estimated by computing the energy along the pathway without any relaxations. This provides an upper bound for the actual barrier. In cases where the upper limit is high, the minimal energy pathways are calculated with relaxations performed via solid-state nudged elastic band (SSNEB) method  \cite{SSNEB} using the mapping result as the starting point. 

Note that this procedure assumes concerted motion of atoms during the phase transformation, and as such does not include interface effects associated with nucleation and growth. Because of this, when evaluating if a phase transition is rapid or sluggish, we follow a broadly adopted semi-quantitative distinction \cite{Trinkle2003}. In short, barriers close to 100 meV/atom or lower would be classified as allowing rapid phase transformations, while those with barriers much above this value would be classified as characteristic of sluggish transformations \footnote{For example, barrier for the concerted diamond-to-graphite transformation is estimated to be $\sim$300 meV/atom.}. The transformation mechanisms and energy profiles from this work are collected into a database of pathways that is provided in Supplementary Information.

Exploration of the relevant crystal phases of the three rocksalt compounds from the first-pinciples random structure sampling shows that their potential energy surfaces share many of the same features, as summarized in Fig.~\ref{fig:random_sample}(a).  For MgO, the rocksalt structure is found for more than half of the random structures, or 1061 times out of 2000 total random structures needed to achieve converged results. The frequency of the next most frequent structure is lower by a factor of $\sim$20. For TaC, rocksalt occurs less frequently, making up about 4 \% of the relaxed random structures, but it is more frequent than the next one by a factor of $\sim$5. For PbTe, the rocksalt structure also occurs $\sim$4 \% of the time (80 out of 2000), and it is about 40 times more frequent than the next one.  

The majority of other structures found in the random sampling have no symmetry at all (s.g.~\#1).
They represent a ``sea'' of local minima, the number of which increases exponentially with the system size \cite{Stillinger_PRE:1999}. Extrapolation to an infinite number of initial random structures would return negligible probabilities of occurrence for all of them. There are also more symmetric structures that exhibit stable frequencies of occurrence as a function of the number of random structures. These occur much less  frequently than rocksalt and are consequently less likely to be realized. They share some common structural features with the rocksalt structure. Most are built from different arrangements of 6-fold coordination polyhedra including face-sharing and/or edge-sharing octahedra and trigonal prisms. 
Structures consisting of these motifs include those with s.g.~\# 160, 166, 187, and 194 in MgO, s.g.~\# 160,166,187, and 194 in TaC, and s.g.~\# 160 and 186 in PbTe. MgO is the only chemistry which exhibits symmetric structures that do not all fall into the above categories. These other structures include the tetrahedrally coordinated zincblende (ZB, s.g.~\#216) and wurtzite (WZ, s.g.~\#186) structures, a hexagonal boron nitride (hBN, s.g.~\# 194), and a buckled hBN (s.g.~\#136) structure, as well as a structure consisting of MgO$_6$ octahedra and MgO$_5$ trigonal bipyramids with long channels running in the c-direction (s.g.~\# 140). 

We now turn to the question of the lifetimes of higher energy, symmetric structures found by the random sampling. If any of them could potentially be realized in spite of their narrow local minima it is important to ask whether they would be long-lived (metastable). 
Interestingly, for all three chemistries, structure mapping combined with SSNEB calculations finds that all symmetric structures identified by the random sampling have low energy barriers ($\lesssim100$ meV/atom) for transformations to the rocksalt ground state. The results are summarized in Fig.~\ref{fig:random_sample}(b) and the mechanisms are provided in Supplementary Information as part of the pathways database. 
For the 6-fold coordinated structures, conversion to rocksalt generally involves a slipping process that transforms trigonal prisms to octahedra and/or changes the orientation of octahedra from face- to edge-sharing.
For the structures with MgO$_x$ polyhedra with $x < 6$, mechanisms also include an increase in coordination from 4 or 5 to 6.

Among structures found in random sampling, the highest barrier for transformations to rocksalt is 127 meV/atom. This is for the TaC structures with s.g.~\#194 and 160. Their predicted transformation mechanisms have the same rate limiting steps because the structure with s.g.~\#160 is an intermediate local minimum along the pathway from s.g.~\#194 to the rocksalt structure. As a result of the relatively low energy barriers for transformation to rocksalt, all symmetric structures from the random sampling are expected to have short lifetimes at ambient conditions. 

These results explain the lack of metastable polymorphs in binary rocksalts. They imply that for all three studied chemistries the local minimum corresponding to the rocksalt structure is 5 to 40 times larger than any other and is hence, statistically much more significant. Additionally, all other, statistically non-negligible local minima would be short-lived as the kinetic barriers are not high enough to ensure kinetic trapping. 

\begin{figure}[t!]
\includegraphics[width=0.98\linewidth]{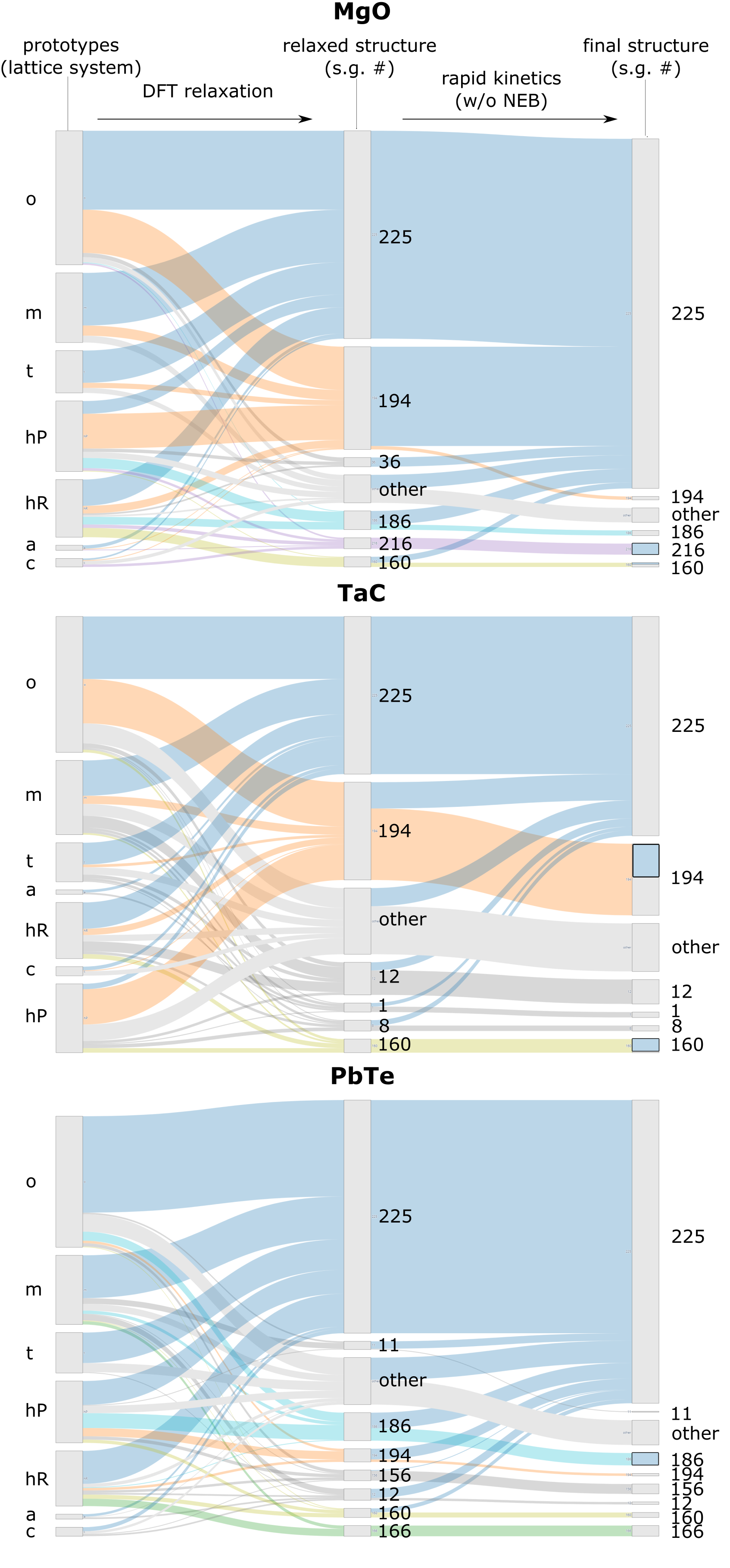}
\caption{\label{fig:proto_path} 
Sankey diagrams showing the kinetic flow of 457 prototype structures found in the ICSD spanning the A$_1$B$_1$ chemical space for each of MgO, TaC, and PbTe. The bars on the left represent the distribution of the  A$_1$B$_1$ prototypes  across 7 crystal systems. Bars in the center display the distribution of space groups that these prototypes relax into. Structures which link to the s.g.~\#225 bar at the right end have kinetic barriers for transformation to rocksalt less than 150 meV/atom. 
}
\end{figure}

{\it What about other structures?}  It is also interesting to investigate whether these conclusions hold for other structures not found by the random sampling. 
According to ICSD, A$_1$B$_1$ compounds crystallize in 457 different structures types (see Methods). To test which of these would constitute local minima for the three studied chemistries and how shallow they might be, we substituted the MgO, PbTe and TaC chemistries in all A$_1$B$_1$ prototype structures from ICSD and relaxed them using DFT. For the resulting relaxed structures, we estimated their kinetics for transformation to the rocksalt using our structure mapping method. The results are summarized in Fig.~\ref{fig:proto_path}.  

\begin{figure*}[t!]
\includegraphics[width=0.8\linewidth]{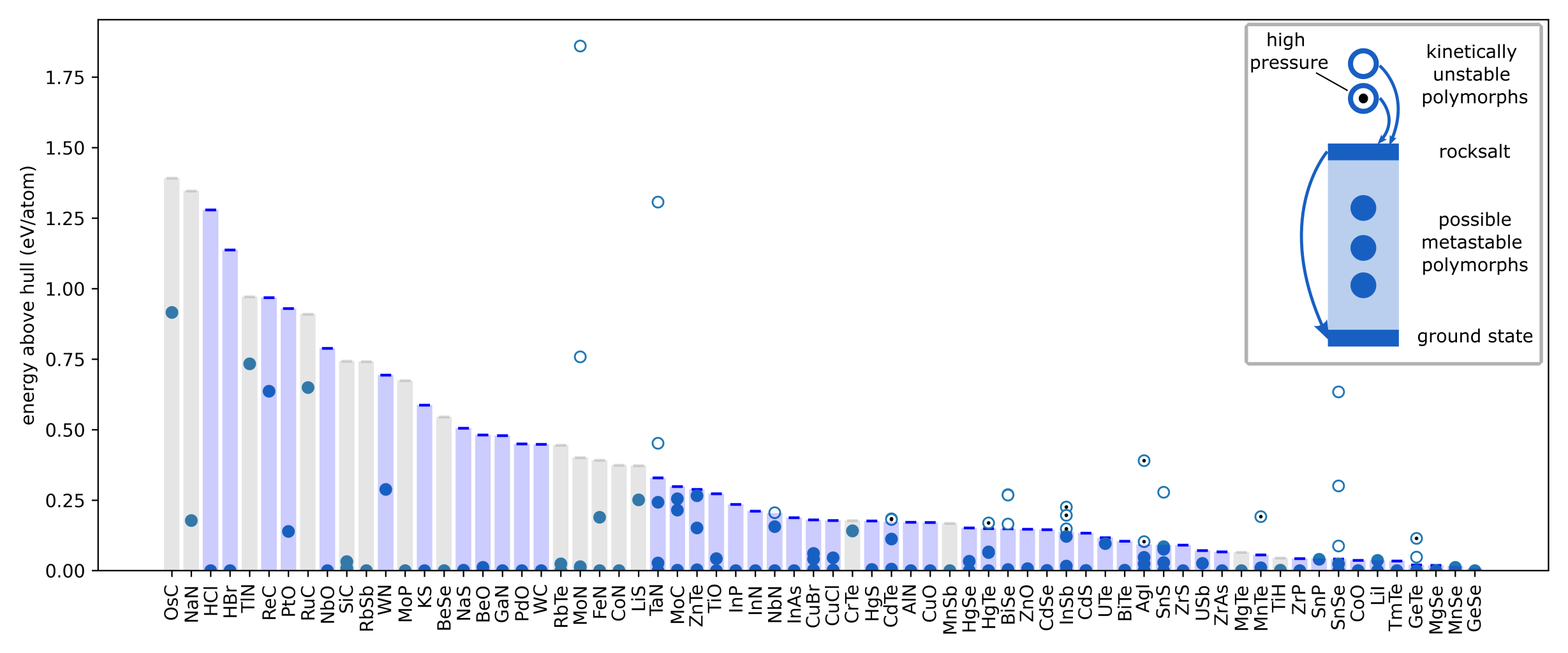}
\caption{\label{fig:limit} 
Energies of reported polymorphs for A$_1$B$_1$ chemistries where rocksalt is reported as a high energy polymorph. Bars represent the energy of the rocksalt relative to the convex hull, experimentally realized rocksalts are blue, while theoretical ones are grey. Other experimentally realized polymorphs with energy below rocksalt are filled blue dots, while polymorphs with energy above rocksalt are open blue circles, and high pressure structures with energy above rocksalt are additionally marked with a black dot. The only reported polymorphs with energy above rocksalt are those in the exception categories specified in the text.}
\end{figure*}

For MgO, TaC, and PbTe, respectively, 245, 174, and 283 of the 457 prototypes relax spontaneously to rocksalt structure implying zero or nearly zero barriers. For the remaining structures, pathways that have static energy profiles with maximum energy below 150 meV/atom are considered characteristic of a rapid transformation to rocksalt.  We use 150 meV/atom because our structure mapping result is an estimate of the upper bound for the barriers. The ssNEB calculations will, in most cases, relax to a minimal energy pathway with a lower barrier. Adding this criterion, a total of 412, 251, and 369 of the 457 prototypes are estimated to transform quickly to the rocksalt structure in MgO, TaC, and PbTe. 

These are relatively large fractions of common A$_1$B$_1$ structures, in particular for MgO and PbTe. TaC seems a little different, but it is important to note that the 108 prototypes that relax directly to structures with s.g.~\#194 include 38 that end in the nickeline (NC)  
structure type, which has a low barrier for transformation into rocksalt according to our SSNEB calculations (see Fig.~\ref{fig:random_sample}(b)). These are marked by the blue rectangle in Fig.~\ref{fig:proto_path}. The same is true for s.g.~\#160 in TaC,  s.g.~\#216 for MgO, and s.g.~\#186 for PbTe. In summary we can say that 291 out of 457 prototypes (64\%) transform quickly to rocksalt in TaC.
While not all prototypes transform rapidly into rocksalt, those that do not are, according to the random structure sampling, statistically insignificant and based on the collected evidence should be very hard if not impossible to realize. 
 
{\it What if rocksalt is not the ground state?} For many A$_1$B$_1$ compounds the ground state is not rocksalt, but it does appear as either a high pressure phase or a phase that can be epitaxially stabilized. Our results are relevant for these situations too, as only the structures that have energies in between the ground state and the rocksalt structure are plausible candidates for the metastable states. Most of those with energies above rocksalt will tend to transform rapidly into the rocksalt which will then likely transform quickly into the ground state, making them short-lived. In this way, the energy of the rocksalt structure then represents an energy limit for the likely metastable states. This is of course true for the larger fraction of prototypes that do transform rapidly. To what extent the rocksalt structure a limit of metastability for the A$_1$B$_1$ chemistry can be seen in Fig.~\ref{fig:limit}.  

We find that for most A$_1$B$_1$ compositions which have rocksalt as a high energy structure according to MP \cite{materialsproject}, other experimentally reported polymorphs have their total energy equal or below the rocksalt structure. 
The only exceptions to this trend (open circles) are (i) high pressure phases that transform quickly to rocksalt and are not metastable (open circles with black dot), (ii) misfit layered compounds, which are formed  by thin layers of two different compounds (see SI) \cite{Wiegers1996}, and (iii) instances with misassigned structure type.
This is all evident if we take SiC as an example. It has a relatively corrugated potential energy surface with polytypes in the family of zincblende, wurtzite, and a tetrahedrally coordinated rhombohedral structure all forming as metastable polymorphs and all having energies below rocksalt as confirmed by the random structure sampling \cite{polysamplercarbides}. A polymorph network and a Sankey diagram for SiC are provided in Supplementary Figures \ref{SiC_polymorph_network} and \ref{SiC_sankey}. 

{\it Metastable polymorph discovery beyond A$_1$B$_1$.} The realization that in A$_1$B$_1$ chemistries most structures with energies above rocksalt will transform quickly to rocksalt and then to the respective ground-state may offer a general insight for the discovery of metastable polymorphs in other compositions. Stoichiometries other than A$_1$B$_1$ may behave the same way provided that a structure playing the role of the rocksalt exists.
Some evidence for the A$_1$B$_2$ chemistry can be found in the literature. Namely, the fluorite structure (also s.g.~\# 225) is the most commonly observed A$_1$B$_2$ structure and it is well documented (Ref.~\cite{pmpaths} and the references therein) that the transformations from the highest-pressure fluorite structure to other observed lower pressure phases of SnO$_2$ including the ground-state rutile are all rapid, indicating fluorite may play a role similar to that of rocksalt in A$_1$B$_1$ (See Supplementary Information Table \ref{fluorite_polymorphs}). Indeed, A$_1$B$_2$ compositions with the fluorite ground state have only cotunnite as a possible metastable polymorph, and there are reports of it also transforming rapidly into fluorite \cite{Morris2001}. Additionally, all known polymorphs of SiO$_2$ have their total energies below that of fluorite (see Supplementary Information Table \ref{SiO2_E}).

{\it Relevance to high entropy systems.} The arguments connecting ``widths'' of the local minima on the potential energy surface to experimental realizability 
apply also to atomic disorder and entropy stabilized systems. An intriguing fact is that virtually all equimolar mixtures of A$_1$B$_1$ binaries are stabilized in the rocksalt structure even when a number of components are not rocksalts themselves. The exceptions are high-entropy borides, which form in a hexagonal structure with layers of edge sharing octahedra which share faces with the octahedra in adjacent layers, likely due to the tendency of boron to form rings. While simplistic entropy arguments based on counting configurations of atoms on a given lattice 
offer an explaination for the formation of the mixture as opposed to the phase separation \cite{CurtaroloNatureCarbides,NavrotskyHEOThermo}, there is no explanation as to why the rocksalt structure is observed and not some other. Namely, contributions to configurational entropy from different arrangements of atoms are the same irrespective of the type of the crystal structure. 

Based on the previous discussion we can say that in a given mixture each individual arrangement of atoms having the rocksalt structure as the underlying (parent) structure could be considered a relatively ``wide'' local minimum. The width of the rocksalt local minimum is also relevant for mixtures as it allows relatively large distortions away from the ideal rocksalt structure while still remaining inside the rocksalt basin of attraction. Then all those arrangements of atoms (cations and/or anions) will collectively form a very wide ``valley'' on the potential energy surface corresponding to the disordered rocksalt phase. Because of all this it could be hypothesized that rocksalt will remain statistically the most significant structure even when a large fraction of components are not rocksalts themselves.

We put this hypothesis to test by performing the random structure sampling for a hypothetical (MgO)(ZnO)(CuO) mixture. Out of the three oxides, only MgO crystalizes in the rocksalt structure, while both ZnO and CuO strongly favor 4-fold coordination in their ground states. Random structure sampling results show disordered rocksalt as the highest occurring structure-type with different atomic configurations spanning a relatively narrow energy range as evidenced by the thermodynamic density of states (see Fig.~\ref{MgZnCuO3} in Supplementary Information). 
Hence, in spite of only 1/3 of cations favoring rocksalt structure, it remains statistically the most significant crystal structure of all. 

Another interesting example supporting our discussion is ZnZrN$_2$ ternary nitride, which is found in thin-film synthesis to always form in the disordered rocksalt structure in spite of having a well-defined non-rocksalt (layered) ground state \cite{ZnZrN2}. The case of ZnZrN$_2$ is particularly interesting as many other structure-types are calculated to have lower energies than the cation-disordered rocksalt phase \cite{ZnZrN2}. The random structure sampling reported in Ref.~\cite{ZnZrN2} once again shows that the cumulative frequency of occurrence of the collection of the disordered rocksalt configurations is $\sim$100 times higher than the ground state, which appears as a  very narrow local minimum in comparison to the rocksalt. Hence, our findings are also consistent with the difficult experimental realization of the ZnZrN$_2$ in its ordered ground state structure and the realization of the disordered rocksalt instead, as a consequence of the high configurational entropy of the disordered rocksalt originating from the large fraction of the configurational space that can be assigned to the rocksalt structure.

\section{Conclusion}
In this paper we examine the lack of metastable polymorphs among compounds with A$_1$B$_1$ stoichiometry and the rocksalt ground state. Using first-principles random structure sampling and transformation kinetics modeling we show that a combination of probabilistic and kinetic factors is responsible for the lack of polymorphism in this important group of materials. Namely, we demonstrate using ionic MgO, metallic TaC and covalent PbTe as case studies, that metastable polymorphs are not found among A$_1$B$_1$ rocksalts because (a) the rocksalt structure represents a very large local minimum on the potential energy surface in comparison to other structures, which makes it statistically much more significant than any other, and (b) that virtually all relevant crystal structures including those found by the random structure sampling as well as the majority of the A$_1$B$_1$ structure prototypes transform rapidly to the rocksalt structure. Moreover, the minority of the prototypes that do not transform rapidly all represent a very narrow local minima according to the random sampling, and, are therefore statistically insignificant. In other words, while they may be candidate metastable structures their experimental realization is predicted to be very difficult if not entirely unlikely.  In addition, we showed that same arguments can be extended to explain observed polymorphs in A$_1$B$_1$ compounds that do not have rocksalt ground state, and possibly also in other chemistries. Probabilistic factors (``width'' of local minima) are also found to be responsible for the realization of the disordered rocksalt structure in high-entropy ceramics and disordered ternary nitrides. Lastly, all  transformation pathways from A$_1$B$_1$ prototypes to the rocksalt structure are collected in a database provided in Supplementary Information.

\begin{acknowledgments}
This work is supported by the National Science Foundation, Grant No. DMR-1945010 and was performed using computational resources provided by Colorado School of Mines and using services provided by the PATh Facility, which is supported by the National Science Foundation award \#1836650. The support of the NSF-funded REU program at Colorado School of Mines, Grant No. DMR-1950924, is also acknowledged.
\end{acknowledgments}

\section{Methods}
\subsection{First-principles random structure sampling}
We evaluate which local minima are statistically relevant via the first-principles random structure sampling \cite{randomsuperlattice}. In this sampling, lattice vectors with a random choice of parameters $a,b,c,\alpha,\beta,\gamma$ are generated. A given number of atoms are then distributed over two interpenetrating grids of points, one for cations and the other for anions, to promote cation-anion coordination. Grids are constructed by discretizing a family of planes defined by the reciprocal lattice vector ${\bf G}=n_1{\bf g}_1+n_2{\bf g}_2+n_3{\bf g}_3$ with a random choice of $n_1,n_2,n_3$. Cations and anions are distributed along the minima and maxima of a plane wave defined by $cos({\bf Gr})$, respectively. To prevent atoms from being placed too close to one another, a probability distribution is constructed of Gaussians centered at the location of each atom, and new atoms are placed in regions where the sum of Gaussians is low. In this work supercells with 24 atoms are generated, and the number of random structures is chosen based on the convergence of the appearance of different structures within the samples. Plots demonstrating this convergence can be found in the Supplementary Information Figure \ref{poly_sampler_conv}. This results in approximately 2000 structures for MgO, 1000 structures for TaC, and 2000 structures for PbTe. Once a structure is generated, it is then relaxed using DFT to its closest local minimum on the potential energy surface. After relaxations are finished the relaxed structures are then grouped into classes of equivalent structures if (a) their space group assignments are the same, (b) their energies are within 10 meV/atom, (c) their volumes are within 5 \%, and (d) and the coordination numbers of atoms up to 3rd coordination shell are the same.

\subsection{Density functional theory relaxations}
Relaxations of the random structures include all degrees of freedom (cell shape, volume, atomic positions) and are performed using the conjugate gradient algorithm such that each structure relaxes to its closest local minimum on the potential energy surface with high probability. Forces for the conjugate gradient algorithm are evaluated with density functional theory \cite{DFT} using the Vienna {\it ab initio} Simulation Package (VASP) \cite{VASP1}. The relaxtions are performed with the generalized gradient approximation, specifically, the Perdew-Burke-Ernzerhof exchange-correlation functional \cite{GGA}. Electronic degrees of freedom are described within the projector-augmented wave (PAW) formalism \cite{paw}. The plane-wave cutoff used in all calculations is 540 eV, and the Monkhorst-Pack {\bf k}-point grids \cite{mp_PRB:1976} are automatically generated with a length parameter of 20 to achieve uniform subdivisions along the reciprocal lattice vectors in all structures. The electronic structure is considered converged when the change in total energy between steps of the Davidson algorithm is less than $10^{-6}$ eV. The ionic positions are considered converged when the change in energy between steps of the conjugate gradient algorithm for ionic positions is less than $10^{-5}$ eV. For numerical reasons cell shape and volume relaxations are restarted at least 3 times followed by a self-consistent run. All structures with final pressures above 3 kbar calculated in the final, self-consistent run are re-relaxed until final pressures are all below 3 kbar.

\subsection{Barrier approximation}
Once high symmetry structures are identified via random sampling, their barrier for a concerted transformation to the rocksalt structure is approximated. Optimal atom-to-atom mapping is performed between each relevant local minimum and the rocksalt structure as in \cite{pmpaths}. In this mapping, different representations of the unit cell of each structure are compared to identify those combinations with maximum volumetric overlap, to minimize the distortion of the unit cell along the pathway. For cells with maximum volumetric overlap, atoms are matched between the two structures by way of the Munkres algorithm \cite{munkres}, an iterative solution of the assignment problem. Iterating over unit cell combinations with the Munkres algorithm returns mappings with minimized distance traveled by the atoms during the concerted transformation. For each mapping, the first shell coordination is computed for ten images along the transformation. We find that mappings with monotonic changes in coordination have smaller barriers, so these are used as starting points where possible. If there is no path with monotonic change in coordination, the path chosen is that with the smallest change in coordination, if multiple paths have the same change in coordination along the path, the path with the shortest distance is used. 

An upper bound for the energy barrier of this transformation can be computed by calculating the energy of each image along the path. Energy is calculated using the same parameters as the relaxations, but the structures along the path are not allowed to relax. The energies are calculated for structures identified from the random sample, if the barrier ($E_{max,path}-E_{polymorph}$) is found to be greater than 100 meV/atom, we perform solid state nudged elastic band (SSNEB) calculations to find the minimum energy pathway near the pathway predicted as above \cite{SSNEB}. The SSNEB calculations are considered converged when the force perpendicular to the tangent along the path is less than 0.1 eV/Angstrom. Initial calculations use 9 intermediate images, if these calculations exhibit intermediate local minima, the local minima are relaxed, and a new SS-NEB calculation is performed between the local minima along the pathway in order to converge each step along the path. 

\subsection{Database of transformation mechanisms}
To determine if prototypical A$_1$B$_1$ structures are likely to transform quickly to the rocksalt structure and to begin the population of a database of NEB starting points, we used the AFLOW Xtalfinder \cite{xtalfinder} to classify each A$_1$B$_1$ structure in the ICSD as belonging to an AFLOW prototype \cite{aflowprototype}. For some prototypes, all atoms occupy fixed Wyckoff positions and the only variable parameter is the lattice constant. For these cases, only a single representative of that prototype is needed in each chemistry. For those where there exist structures in the ICSD with different lattice constant ratios and variable wyckoff positions such that atom positions vary between members of the same prototype, structures are classified as equivalent using the AFLOW Xtalfinder \cite{xtalfinder}, such that every distinct structure found in the ICSD for a given prototype is represented. If swapping A and B sites leads to different coordination of A and/or B atoms, the swapped structure is also considered. 

For each prototypical structure identified in this way, the A and B sites are populated with the elements of the three example compounds. The lattice constant is then scaled such that the average bond length between A and B atoms is equal to the bond length found in the rocksalt structure for each chemistry. Atoms are then perturbed by 0.3 Angstroms, and the perturbed structures are relaxed as described in the ``Density Functional Theory Calculations'' section. 

The resulting structures are grouped according to their AFLOW prototype labels, and the resulting distribution of space groups is the second column of nodes in Fig.~\ref{fig:proto_path}. An atom-to-atom map is generated from each unique structure to the rocksalt structure, as in the ``Barrier approximation'' section. Only the static barriers are computed for these structures. Those structures with a kinetic barrier to the rocksalt structure of 150 meV or less are considered to have rapid kinetics, and are added to the rocksalt node in the third column of Fig.~\ref{fig:proto_path}. The structures with static barriers of greater than 150 meV/atom are grouped by their space group in third column of Fig.~\ref{fig:proto_path}.

%

\clearpage 

\title{All ``roads'' lead to rocksalt structure - Supplementary Information}

\author{Matthew Jankousky}                                                                                                     
\affiliation{Colorado School of Mines, Golden, CO 80401, USA}

\author{Helen Chen}
\affiliation{Harvey Mudd College, Claremont, CA 91711, USA}

\author{Andrew Novick}                                                                                                     
\affiliation{Colorado School of Mines, Golden, CO 80401, USA}

\author{Vladan Stevanovi\'{c}}
\email{vstevano@mines.edu}                                                                     
\affiliation{Colorado School of Mines, Golden, CO 80401, USA}

\date{\today}

\maketitle

\begin{longtable*}{l c l c l c l c l}
\toprule
 compound &  polymorph space groups &     structure types &     realization condition &                                          citations \\
\hline
     INa &                     141 &                     &             mislabeled              &                  \cite{Swanson1955,Kirkpatrick1927} \\
     PrSb &                     123 &                CsCl &             high pressure &                    \cite{Samsonov1974,Hayashi2000} \\
     CeSb &                     123 &                CsCl &             high pressure &                      \cite{Holbourn1981,Leger1984} \\
     PuSb &                123, 221 &                CsCl &             high pressure &  \cite{Dabos-Seignon1990,Dabos-Seignon1990,Makode2013} \\
      IRb &                     221 &                CsCl &             high pressure &                        \cite{Posnjak1922,Weir1964} \\
    LaAs &                     123 &                CsCl &             high pressure &                      \cite{Ugur2008,Shirotani2003} \\
    PbTe &                 221, 62 &  CsCl, Distorted RS &           high pressure &  \cite{Kabalkina1968,Li2013,Korkosz2014,Fujii1984} \\
      KCl &                     221 &                CsCl &             high pressure &                         \cite{Walker2004,Weir1964} \\
      TlI &                 221, 63 &  CsCl, Distorted RS &           high pressure &      \cite{Shamovskii1968,Blackman1961,Becker2004} \\
      CTl &                 99, 221 &                CsCl &             high pressure &         \cite{Penneman1958,Strada1934,Wilhelm1950} \\
      MgO &                186, 187 &            Wurtzite &                 thin film &  \cite{Johnsen2009,Tsirel'son1998,Zwijnenburg2011} \\
     CaTe &                     221 &                CsCl &             high pressure &                         \cite{Zimmer1985,Sifi2012} \\
     TlCl &                 221, 63 &  CsCl, Distorted RS &           high pressure &       \cite{Blackman1961,Ungelenk1962,Roberts2006} \\
     ClAg &                  11, 63 &   Distorted RS, KOH &           high pressure &                  \cite{Hull1999,Hull1999,Hull1999} \\
      HCr &                     186 &            Wurtzite &                 H interstitial in metal, theoretical &                     \cite{Snavely1949,Snavely1949} \\
     TeSr &                     221 &                CsCl &             high pressure &                       \cite{Zimmer1985,Zimmer1985} \\
      SMn &           216, 186, 221 &      CsCl, Wurtzite &  high pressure, alloying &  \cite{Corliss1956,Corliss1956,Wiedemeier1969,Mehmed1938} \\
     SmTe &                     221 &                CsCl &             high pressure &                \cite{Chatterjee1972,Yarembash1971} \\
    NpAs &                     221 &                CsCl &             high pressure &                     \cite{Langridge1994,Dabos1986} \\
     PbSe &         42, 221, 62, 63 &  CsCl, Distorted RS &           high pressure &  \cite{Wang2015,Wang2015,Chattopadhyay1986,Mariano1967,Auriel1993} \\
     TePr &                     221 &                CsCl &             high pressure &               \cite{Chatterjee1972,Chatterjee1972} \\
     SnTe &                216, 221 &                CsCl &             high pressure &   \cite{Goldschmidt1927,Onodera1986,Rogacheva1986} \\
     BrCs &                     221 &                CsCl &             high pressure &                    \cite{Blackman1961,Posnjak1922} \\
     BaTe &                     221 &                CsCl &             high pressure &               \cite{Grzybowski1984,Grzybowski1971} \\
     SeBa &                     221 &                CsCl &             high pressure &                  \cite{Lyskova1975,Grzybowski1983} \\
       KN &                     129 &                     &                     mislabeled      &                          \cite{Juza1957,Nagib1977} \\
      OBa &                129, 194 &                NiAs &             high pressure &            \cite{LiuLingu1971,Taylor1984,Weir1986} \\
     ThAs &                     221 &                CsCl &             high pressure &                       \cite{Gerward1988,Dabos1986} \\
     NdSb &                     123 &                CsCl &             high pressure &                 \cite{Manfrinetti2009,Hayashi2000} \\
     TbSe &                     186 &            Wurtzite &                 thin film &                       \cite{Eliseev1987,Singh1977} \\
     SmSe &                     186 &            Wurtzite &                 thin film &                        \cite{Miller1962,Singh1977} \\
     PrAs &                     123 &                CsCl &             high pressure &                  \cite{Shirotani2001,Iandelli1937} \\
      SSr &                     221 &                CsCl &             high pressure &                        \cite{Flahaut1962,Ugur2009} \\
     BrRb &                     221 &                CsCl &             high pressure &                        \cite{Weir1964,Cortona1992} \\
      NAg &                      64 &                     &              mislabeled             &                         \cite{Chen2010,Niggli1959} \\
     PuAs &                     221 &                CsCl &             high pressure &            \cite{Dabos-Seignon1989,Charvillat1973} \\
      PCe &                     221 &                CsCl &             high pressure &                           \cite{Vedel1987,Ono1974} \\
    LaN &                     129 &           CsCl          &         high pressure      &            \cite{Schneider2012,GiorgioL.Olces1979} \\
      HCs &                     221 &                CsCl &             high pressure &                  \cite{Ponyatovskii1985,Zintl1931} \\
     TeIn &                140, 221 &                CsCl &             high pressure &       \cite{Faita2011,Faita2011,Chattopadhyay1985} \\
     DySe &                     186 &            Wurtzite &                 alloyed thin film &                        \cite{Singh1977,Olcese1961} \\
      LaS &                 64, 129 &            CsCl         &         misfit layered compound, high pressure                 &           \cite{Friese1997,deBoer1991,Charifi2012} \\
     SnSb &                216, 221 &                CsCl &             high pressure &   \cite{Goldschmidt1927,Kolobyanina1972,Losev1970} \\
     ClNa &                     221 &                CsCl &             high pressure &                      \cite{Abrahams1965,Zhang2013} \\
      SPb &  5, 39, 12, 221, 62, 63 &  CsCl, Distorted RS &           high pressure &  \cite{Onoda1990,Wiegers1990,Meerschaut1992,SakthiSudarSaravanan2015,Chattopadhyay1986,Kabalkina1968,Zagorac2011,vanSmaalen1991} \\
     EuTe &                     221 &                CsCl &             high pressure &               \cite{Chatterjee1972,Chatterjee1972} \\
     LaSe &                      12 &                     &          misfit layered compound                &                         \cite{Ren1997,Charifi2012} \\
      SmS &                     139 &        Distorted RS &          misfit layered compound                 &                 \cite{Meerschaut1991} \\ 
     NpTe &                     221 &                CsCl &             high pressure &                \cite{Wastin1995,Dabos-Seignon1990} \\
      MnO &                     186 &            Wurtzite &                 thin film &                    \cite{NamKiMi2012} \\ 
      AsU &                     221 &                CsCl &             high pressure &                 \cite{Trzebiatowski1964,Dabos1986} \\
     NPr &                     129 &                  CsCl   &      high pressure                     &                       \cite{Ettmayer1979,Cynn2010} \\
      CaO &                      64 &                     &          composite crystal                 &                       \cite{Isobe2007,Oftedal1927} \\
     LaSb &                     123 &                CsCl &             high pressure &                      \cite{Leger1984,Holbourn1981} \\
      FCs &                     221 &                CsCl &             high pressure &                        \cite{Weir1964,Posnjak1922} \\
     BrAg &                      11 &                 KOH &             high pressure &                           \cite{Hull1999,Hull1999} \\
       KI &                     221 &                CsCl &             high pressure &                   \cite{Vegard1921,Piermarini1962} \\
     BrTl &                 221, 63 &  CsCl, Distorted RS &           high pressure &         \cite{Singh2011,Ungelenk2004,Blackman1961} \\
     NdAs &                     123 &                CsCl &             high pressure &                    \cite{Shirotani2001,Taylor1974} \\
    ClCs &                     221 &                CsCl &             high pressure &                        \cite{West1934,Swanson1953} \\
      FRb &                     221 &                CsCl &             high pressure &              \cite{Piermarini1962,Goldschmidt1926} \\
     PuTe &                     221 &                CsCl &             high pressure &                \cite{Dabos-Seignon1990,Kruger1967} \\
     SmAs &                     123 &                CsCl &             high pressure &                    \cite{Shirotani2001,Beeken1981} \\
     NpSb &                     123 &                CsCl &             high pressure &              \cite{Dabos-Seignon1990,Mitchell1971} \\
      KBr &                     221 &                CsCl &             high pressure &                       \cite{Dewaele2012,Finch1936} \\
     ClRb &                     221 &                CsCl &             high pressure &                            \cite{Ott1926,Weir1964} \\
     ThSe &                     221 &                CsCl &             high pressure &                         \cite{Olsen1988,d'Eye1952} \\
      ICs &                 221, 62 &  CsCl, Distorted RS &           high pressure &       \cite{Beintema1937,Posnjak1922,Blackman1961} \\
     BiCe &                123, 221 &                CsCl &             high pressure &         \cite{Feng2012,Feng2012,Abdusalyamova1988} \\
       FK &                     221 &                CsCl &             high pressure &                           \cite{Weir1964,Hull1919} \\
     ThSb &                     221 &                CsCl &             high pressure &                     \cite{Gerward1988,Gerward1988} \\
     EuS &                     139 &        Distorted RS &            misfit layered compound               &                       \cite{Cario1998,Eliseev1974} \\

\caption{\label{RS_polymorphs} All compounds in the Materials Project Database with a rocksalt ground state and an experimentally reported polymorph. Inspection of references shows that there are no metastable polymorphs.}
\end{longtable*}

\begin{table*}
\begin{tabular}{l c l c l c l c l c |}
\toprule
 space group &  frequency &     $\Delta E_f$ &     NEB barrier  &                                          description  & name \\
 number  &  & (eV/atom) & (eV/atom) & &  \\
  \hline
  {\bf 225} & 1061 & 0 & 0 & rocksalt (RS) & MgO\_24\_1960 \\
  {\bf 194} & 42 & 0.035 & 0.057 & hexagonal boron nitride (hBN) & MgO\_24\_1015 \\
  72 & 1 & 0.060 & & hBN with fault between rings & MgO\_24\_648 \\
  {\bf 186} & 8 & 0.068 & 0.088 & wurtzite (WZ) & MgO\_24\_1977\\
  {\bf 166} & 2 & 0.076 & 0.099 & RS with stacking fault (face sharing octahedra) & MgO\_24\_1988 \\
  {\bf 194} & 1 & 0.076 & 0.030 & RS with stacking fault (face sharing octahedra) & MgO\_24\_345\\
  14 & 1 & 0.078 & & hBN with fault & MgO\_24\_265 \\
  13 & 1 & 0.079 & & distorted trigonal bipyramids & MgO\_24\_202\\
  {\bf 160} & 26 & 0.081 & 0.101 & rhombohedral WZ polytype & MgO\_24\_1296\\
  {\bf 140} & 1 & 0.082 & 0.001 & trigonal bipyramids and octahedra with pores in the c-direction & MgO\_24\_141 \\
  36 & 3 & 0.087 & & buckled hBN & MgO\_24\_828 \\
  {\bf 136} & 1 & 0.088 & 0.092 & buckled hBN & MgO\_24\_607 \\
  {\bf 160} & 1 & 0.098 & 0.053 & face and edge sharing octahedra, trigonal prisms & MgO\_24\_1011 \\
  {\bf 166} & 58 & 0.101 &  0.041 & RS with stacking fault (face sharing octahedra) & MgO\_24\_266 \\
  12 & 1 & 0.102 & & RS with channels in c-direction & MgO\_24\_527 \\
  {\bf 216} & 13 & 0.102 & 0.044 & Zincblende (ZB) & MgO\_24\_775 \\
  12 & 1 & 0.105 & &  distorted hBN & MgO\_24\_1806 \\
  36 & 1 & 0.113 & & buckled hBN & MgO\_24\_1699 \\
  {\bf 187} & 8 & 0.130 & 0.30 & face sharing octahedra and trigonal prisms & MgO\_24\_1220 \\
  12 & 1 & 0.130 & & RS with channels in c-direction & MgO\_24\_1971\\
  {\bf 194} & 13 & 0.150 & 0.062 & RS stacking fault - all octahedra face sharing & MgO\_24\_728 \\
  36 & 21 & 0.155 & & square pyramids & MgO\_24\_1080 \\
  12 & 1 & 0.161 & & see-saws and tetrahedra with channels & MgO\_24\_1181\\
  36 & 1 & 0.164 & & square pyramids & MgO\_24\_1817 \\
  38 & 1 & 0.168 & & square pyramids and tetrahedra & MgO\_24\_1246 \\
  {\bf 161} & 1 & 0.218 & 0.009 & disordered 7-fold coordination & MgO\_24\_1609 \\
  43 & 4 & 0.220 & & disordered 5-, 6-, and 7-fold & MgO\_24\_304\\
  15 & 2 & 0.221 & & disordered see-saw, tetrahedra, square pyramids & MgO\_24\_937 \\
  {\bf 194} & 1 & 0.244 & 0.019 & nickeline (NC) & MgO\_24\_1249\\
\end{tabular}
\caption{\label{MgO_random_sample} Random sampling structures identified for MgO}
\end{table*}

\begin{table*}
\begin{tabular}{l c l c l c l c l c |}
\toprule
 space group &  frequency &     $\Delta E_f$ &     NEB barrier  &                                          description  & name \\
 number  &  & (eV/atom) & (eV/atom) & &  \\
  \hline
  {\bf 225} & 45 & 0 & 0 & rocksalt (RS) & C1Ta1\_24\_18 \\
  {\bf 166} & 4 & 0.066 & 0.89 & trigonal prisms and octahedra & C1Ta1\_24\_990 \\
  {\bf 160} & 10 & 0.072 & 0.127 & trigonal prisms and octahedra & C1Ta1\_24\_849 \\
  {\bf 187} & 2 & 0.092 & 0.067 & trigonal prisms and octahedra & C1Ta1\_24\_719 \\
  {\bf 194} & 2 & 0.123 & 0.127 & nickeline (NC) & C1Ta1\_24\_770 \\
  {\bf 166} & 1 & 0.158 & 0.0636 & RS with stacking fault & C1Ta1\_24\_909 \\
  63 & 1 & 0.206 & & distorted hBN & C1Ta1\_24\_451 \\
  12 & 1 & 0.216 & & layerd trigonal prisms, distorted trigonal bipyramids, octahedra & C1Ta1\_24\_345 \\
  44 & 1 & 0.244 & & trigonal prisms and octahedra & C1Ta1\_24\_416 \\
  12 & 1 & 0.264 & & distorted trigonal prisms and 7-fold coordination & C1Ta1\_24\_581\\
\end{tabular}
\caption{\label{TaC_random_sample} Random sampling structures identified for TaC}
\end{table*}

\begin{table*}
\begin{tabular}{l c l c l c l c l c |}
\toprule
 space group &  frequency &     $\Delta E_f$ &     NEB barrier  &                                          description  & name \\
 number  &  & (eV/atom) & (eV/atom) & &  \\
  \hline
  {\bf 225} & 84 & 0 & & rocksalt (RS) & PbTe\_24\_761 \\
  12 & 2 & 0.051 & & RS with channels in c-direction & PbTe\_24\_46 \\
  {\bf 186} & 1 & 0.078  & 0.028 & layered face-sharing octahedra & PbTe\_24\_1279 \\
  {\bf 160} & 2 & 0.084 & 0.007 & layered trigonal prisms and octahedra & PbTe\_24\_1527 \\
  12 & 1 & 0.108 & & disordered octahedra and square pyramids & PbTe\_24\_1532 \\
  12 & 1 & 0.117 & & disordered octahedra and square pyramids & PbTe\_24\_1716 \\
  38 & 1 & 0.131 & & RS with site swap & PbTe\_24\_417 \\
  38 & 1 & 0.174 & & RS with site swap & PbTe\_24\_1226 \\
\end{tabular}
\caption{\label{PbTe_random_sample} Random sampling structures identified for PbTe}
\end{table*}

\begin{table*}
\begin{tabular}{l c l c l c l c l c |}
\toprule
 space group &  frequency &     $\Delta E_f$ &     NEB barrier  &                                          description  & name \\
 number  &  & (eV/atom) & (eV/atom) & &  \\
  \hline
 {\bf 216} & 36 & 0 & 0 & zincblende (ZB) & SiC\_12\_1523 \\
 {\bf 160} & 4 & 0 & 0.237 & tetrahedral rhombohedral WZ polytype & SiC\_12\_723 \\
 {\bf 186} & 25 &0.002 & 0.345 & wurtzite (WZ) & SiC\_12\_1769 \\
 36 & 2 & 0.065 & & disordered tetrahedra & SiC\_12\_949 \\
 36 & 1 & 0.131 & & disordered tetrahedra & SiC\_12\_743 \\
 44 & 1 & 0.190 & & tetrahedra mixed with C-C and Si-Si bonds & SiC\_12\_1296 \\
 15 & 3 & 0.324 & & tetrahedra mixed with trigonal planar SiC$_3$ & SiC\_12\_1149 \\
 12 & 1 & 0.364 & & disordered tetrahedra, C-C bonds & SiC\_12\_1882 \\
 152 & 1 & 0.366 & & disordered tetrahedra & SiC\_12\_690 \\
 20 & 1 & 0.403 & & disordered tetrahedra, C-C and Si-Si bonds & SiC\_12\_1245 \\
 12 & 1 & 0.481 & & disordered tetrahedra, C-C and Si-Si bonds & SiC\_12\_885 \\
 38 & 1 & 0.497 & & edge-sharing tetrahedra & SiC\_12\_128 \\
 25 & 1 & 0.524 & & disordered tetrahedra, C-C and Si-Si bonds & SiC\_12\_61 \\
 44 & 1 & 0.566 & & disordered tetrahedra, trigonal planar SiC$_3$, C-C and Si-Si bonds & SiC\_12\_1859 \\
 43 & 1 & 0.566 & & disordered, distorted tetrahedra & SiC\_12\_364 \\
 107 & 1 & 0.709 & & 5-fold, intermediate between ZB an RS & SiC\_12\_1932 \\
 {\bf 225} & 2 & 0.740 & 0.001 & rocskalt (RS) & SiC\_12\_6
\end{tabular}
\caption{\label{SiC_random_sample} Random sampling structures identified for SiC}
\end{table*}

\begin{figure*}[h!]
\includegraphics[width=\linewidth]{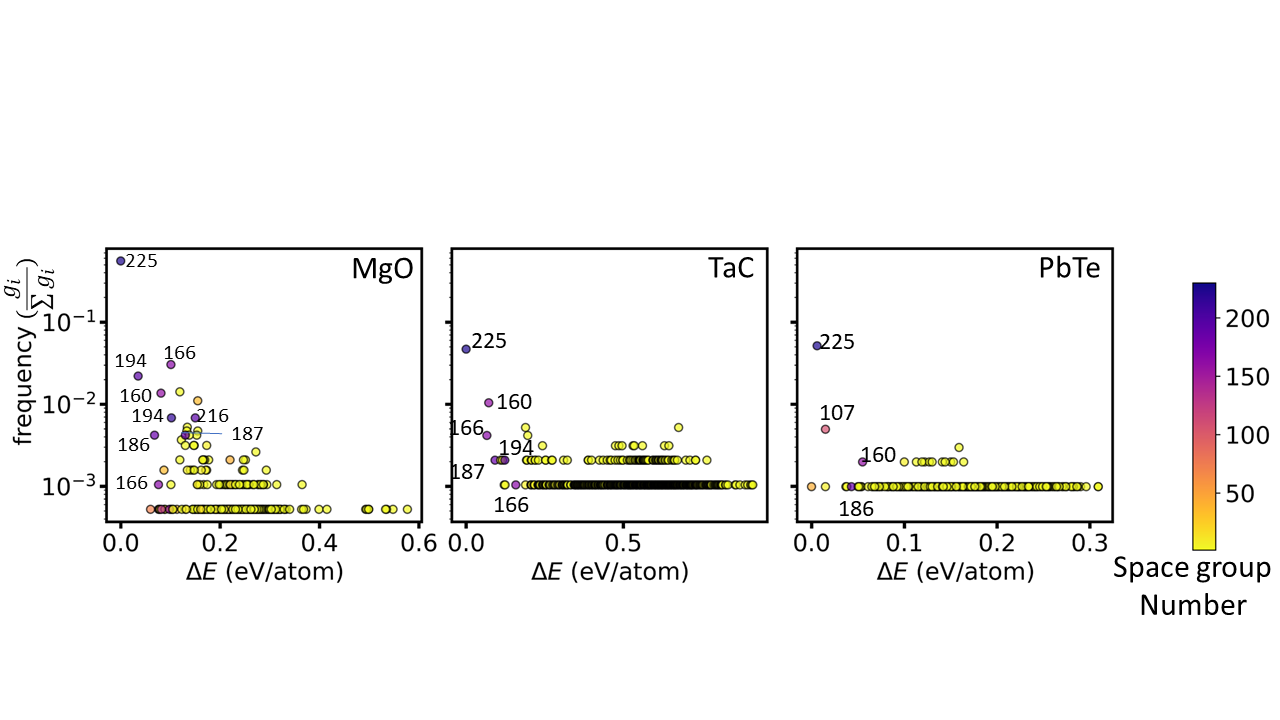}
\caption{\label{RS_frequency_vs_E} The frequency of occurence vs energy for each of MgO, TaC, and PbTe.}
\end{figure*}

\begin{table*}
\begin{tabular}{l c l c l c l c l}
\toprule
 compound &  polymorph space group &     structure types &     realization condition &                                          citations \\
 \hline
 MoN & 191 & layered material & space group misassigned & \cite{Troitskaya1961}\\
 MoN & 74 & 1-D chains & space group misassigned & \cite{ZhaoXuDong2000,Yang2014} \\
 NbN & 194 & Face sharing octahedra & degenerate energy & \cite{Schoenberg1954}\\
 SnSe & 129 & tetragonal layers & misfit layered compound & \cite{Wiegers1991} \\
 SnSe & 63 & 1-D chains & space group misassigned & \cite{Chattopadhyay1986} \\
 SnSe & 59 & 2-D ribbons & space group misassigned & \cite{Chattopadhyay1986} \\
 SnS & 216 & zincblende & Evaporation (citation only) & \cite{Badachhape1962}\\
 SnS & 39 & layered material & misfit layered compound & \cite{Hoistad1995}\\
 SnS & 39 & layered material & misfit layered compound & \cite{Meetsma1989}\\
 SnS & 2 & layered material & misfit layered compound & \cite{Wiegers1992}\\
 CdTe & 63 & distorted RS & high pressure & \cite{Nelmes1995}\\
 CdTe & 63 & distorted RS & high pressure & \cite{Nelmes1995}\\
 GeTe & 62 & Pnma layered & citation only & \cite{Karbanov1968}\\
 GeTe & 60 & buckled square ribbons & high pressure & \cite{Onodera1997}\\
 LiI & 194 & face sharing octahedra & degenerate & \cite{Ott1923}\\
 TaN & 191 & close packed planes with trigonal prisms between & space group misassigned &\cite{Chihi2011}\\
 TaN & 191 & close packed planes with octahedra between & space group misassigned & \cite{Brauer1972} \\
 AgI & 11 & distorted rocksalt & high pressure & \cite{Hull1999} \\
 AgI & 221 & CsCl-type & high pressure & \cite{Adams1962}\\
 HgTe & 63 & distorted rocksalt & high pressure & \cite{HaoAimi2009}\\
 BiSe & 12 & & misfit layered compound & \cite{Clarke2015}\\
 BiSe & 12 & & misfit layered compound & \cite{Clarke2015}\\
 BiSe & 14 & & nonstoichiometric & \cite{Aurivillius1960} \\
 MnTe & 62 & distorted NiAs & high pressure & \cite{Mimasaka1987} \\
 InSb & 59 & distorted Rocksalt & high pressure & \cite{Yu1978}\\
 InSb & 196 & high pressure cadmium telluride & high pressure & \cite{Yu1978}\\
 InSb & 221 & CsCl-type & high pressure & \cite{Vanderborgh1989}
 \end{tabular}
\caption{\label{table:RS_metastable} All compounds in the Materials Project Database with rocksalt reported as a polymorph. Inspection of references shows that there are no metastable polymorphs with energy above that of rocksalt.}
 \end{table*}

\begin{figure*}[h]
\includegraphics[width=0.5\linewidth]{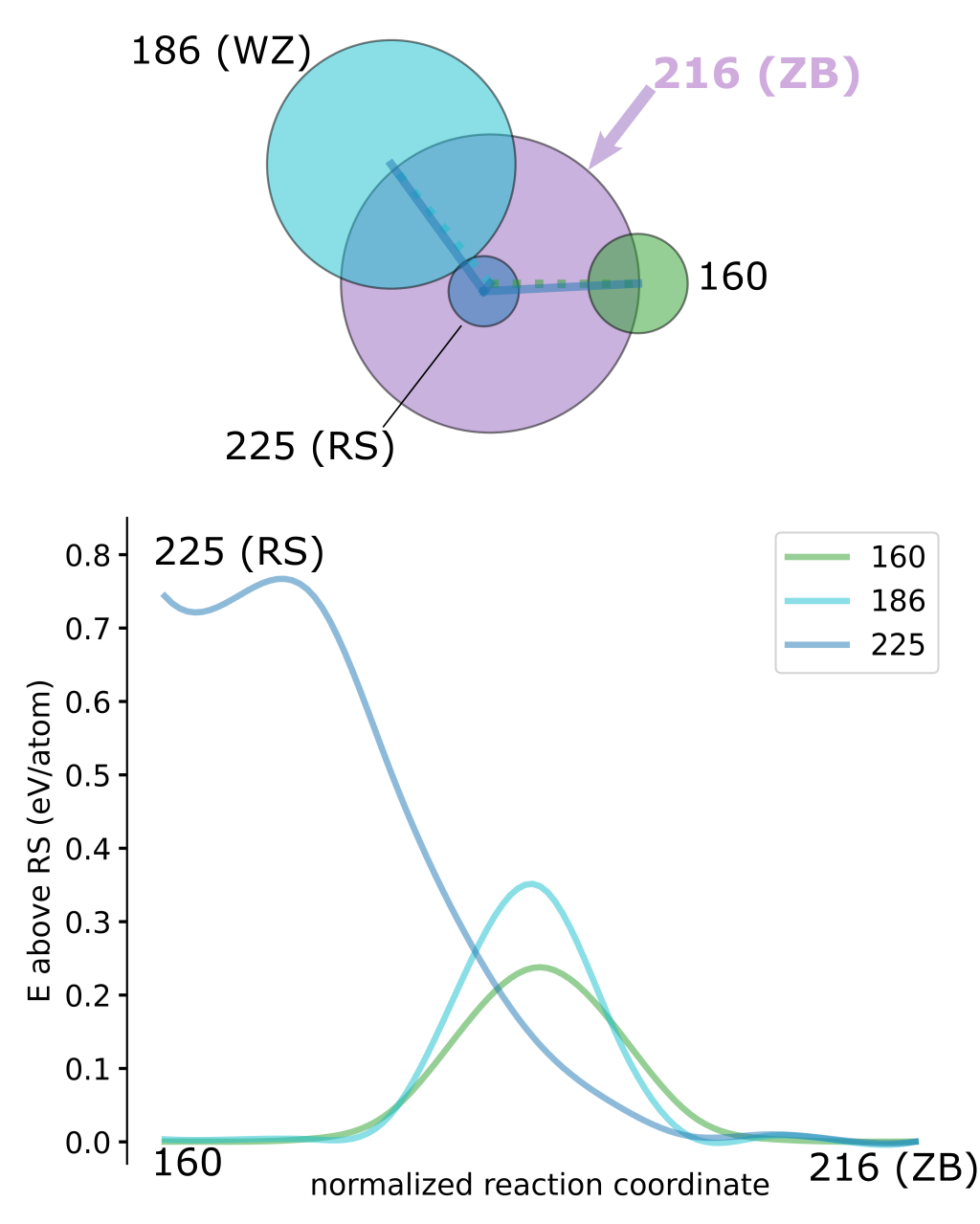}
\caption{\label{SiC_polymorph_network} A polymorph network representing the polymorphism of SiC. Zincblende is taken to be the origin. 
All other structures are labeled according to their space group number.  The size of each node represents the frequency of occurence of a structure in the random sample. Note that a fixed size is used to represent the width of the zincblende local minimum, and other points are scaled relative to this size.
Structures are arranged so that the energy increases with the polar angle and the distance of each node from the zincblende center is proportional to the distance travelled by the atoms during phase transformation of that structure to zincblende.
Dashed lines represent slow transformations from wurtzite and rhombohedral to zincblende. Solid lines represent fast transformations between rocksalt and the tetrahedral polymorphs.}
\end{figure*}

\begin{figure*}[h]
\includegraphics[width=\linewidth]{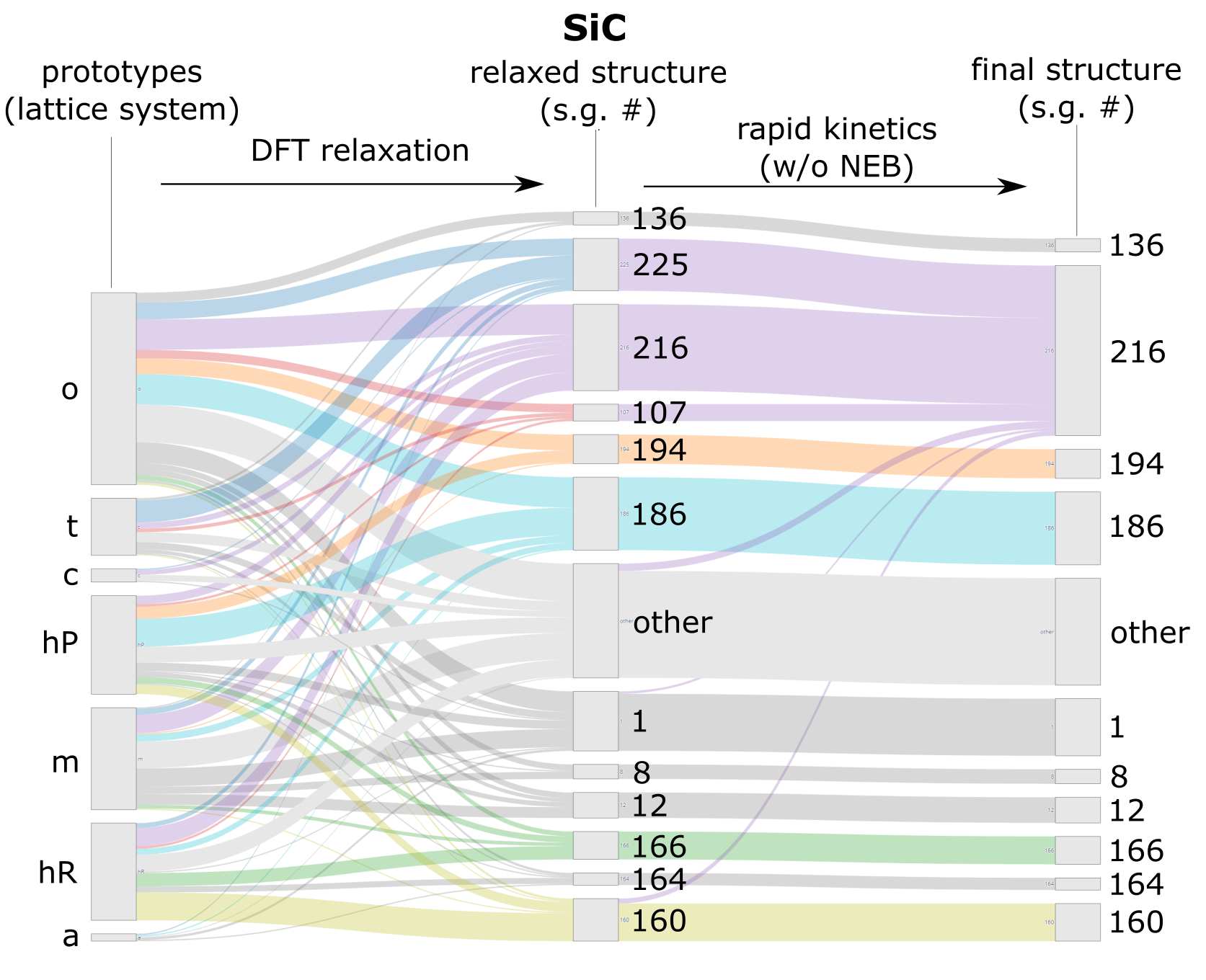}
\caption{\label{SiC_sankey}Sankey diagrams showing the kinetic flow of 457 prototype structures found in the ICSD spanning the A$_1$B$_1$ chemical space for SiC. The bars on the left represent the distribution of the  A$_1$B$_1$ prototypes  across 7 crystal systems. Bars in the center display the distribution of space groups that these prototypes relax into. Structures which link to the s.g.~\#216 bar at the right end have kinetic barriers for transformation to zincblende less than 150 meV/atom. 
For this diagram, we assume that all strucutres with energy greater than the energy of rocksalt will transform quickly to rocksalt, and will therefore transform quickly to the zincblende ground state. Even under this condition, many fewer structures transform to the zincblende ground state in SiC than those that transform quickly to rocksalt in each of the figures in main body. }
\end{figure*}

\begin{table*}
\begin{tabular}{l c l c l c l c l}
\toprule
 compound &  polymorph space groups &     structure types &     realization condition &                                          citations \\
 \hline
     Mg$_2$Sn &                     227,194 &       Laves, cotunnite, hexagonal              &             high pressure, high temperature & \cite{YuFe2011,Boudemagh2011} \\
     K$_2$S & 194 &  & mislabeled (refers to K2SO4 compound) & \cite{Fischmeister1962} \\
     Rb$_2$Te & 227, 194 & cotunnite, Ni$_2$In & high pressure, high temperature      & \cite{Stoewe2004} \\
     Mg$_2$Pb & 62 &  & alloying 
     & \cite{Eldridge1965} \\
     Mg$_2$Si &                     227,194 &       cotunnite, hexagonal              &             high pressure, high temperature          &                  \cite{Boudemagh2011} \\      Cs$_2$Se & 62, 43 & cotunnite, distorted fluorite & cotunnite degenerate, high pressure    & \cite{Sommer1977} \\
     CdF$_2$ & 227 & cotunnite & high pressure (reversible) &  \cite{Liu2011}\\
     Rh$_2$As & 62 & cotunnite & high temperature & \cite{Kjekshus1972} \\
     Mg$_2$Ge & 227 & Cubic Laves phase & high pressure & \cite{LaBotz1963} \\
     Li$_2$S & 62 & anticotunnite & high pressure (reversible) & \cite{Grzechnik2000} \\
     ThO$_2$ & 62 & cotunnite & high pressure (reversible, sluggish) & \cite{Idiri2004} \\
     CaF$_2$ & 62 & cotunnite & high pressure (partially reverts) & \cite{Gerward1992,Morris2001}\\
     SrF$_2$ & 62,194 & cotunnite, Ni$_2$In & high pressure (partially reverts) & \cite{Wang2012} \\
     Rb$_2$S & 62,194 & anticotunnite, Ni$_2$In & high pressure & \cite{Santamaria-Perez2011} \\
     Na$_2$S & 62, 194 & anticotunnite, Ni$_2$In & high pressure (reversible) & \cite{Vegas2001} \\
    	Ba$_2$F & 62, 194 & cotunnite, Ni$_2$In & high pressure (partially reverts) & \cite{Leger1995} \\
    	Ba$_2$Cl & 62 & cotunnite, hydrated structure & high temperature (reversible), hydration & \cite{Hull2011,Haase1978} \\
    	PbF$_2$ & 62 & cotunnite & approximately degenerate & \cite{Ehm2003} \\
\end{tabular}
\caption{\label{fluorite_polymorphs} All compounds in the Materials Project Database with a fluorite ground state and an experimentally reported polymorph. Inspection of references shows that while cotunnite is reported to be metastable for some chemistries, in those cases transformation back to fluorite is at least partially observed at room temperature.}
\end{table*}

\begin{table*}
\begin{tabular}{ c  c  c  c }
\toprule
 ICSD label &  polymorph space groups &     name &     energy (eV/atom) \\
 \hline
091736 & 14 & Cristobalite & -7.908 \\
647410 & 154 & Quartz low & -7.887 \\
190370 & 136 & Stishovite & -7.730 \\
075486 & 92 & Cristobalite low - alpha & -7.904 \\
067121 & 152 & Quartz low & -7.897 \\
077459 & 227 & Cristobalite beta high & -7.905 \\
040900 & 20 & Tridymite & -7.909 \\
089277 & 154 & Quartz low & -7.899 \\
089289 & 180 & Quartz high & -7.899 \\
056473 & 4 & Tridymite & -7.909 \\
162660 & 122 & Cristobalite beta & -7.909 \\
172295 & 15 & Coesite & -7.869 \\
094091 & 19 & Tridymite & -7.908 \\
030795 & 40 & Tridymite O & -7.910 \\
162616 & 198 & Cristobalite beta & -7.905 \\
200478 & 194 & Tridymite 2H high (Gibbs Model) & -7.906 \\
162618 & 122 & Cristobalite beta & -7.910 \\
040137 & 15 &  & -7.902 \\
029343 & 182 & Tridymite 2H low (subcell) & -7.907 \\
192626 & 58 & Ferrierite, siliceous & -7.900 \\
056320 & 227 & Dodecasil 3C & -7.902 \\
067669 & 15 & Moganite & -7.899 \\
064980 &  & Quartz high & -6.860 \\
085586 & 166 & Chabazite & -7.897 \\
048153 & 191 & Dodecasil 1H & -7.905 \\
048154 & 227 & Dodecasil 3C & -7.902 \\
183701 & 229 & Sodalite & -7.904 \\
065497 & 71 &  & -7.900 \\
093975 & 181 & Quartz high & -7.899 \\
            & 225 & Fluorite & -6.661 \\
            &       & Glass ensemble & -7.529 \\
\end{tabular}
\caption{\label{SiO2_E} Energies of named polymorphs of SiO$_2$. All polymorphs have energy below that of SiO$_2$ in the fluorite structure.}
\end{table*}

\begin{figure*}[ht!]
\includegraphics[width=\linewidth]{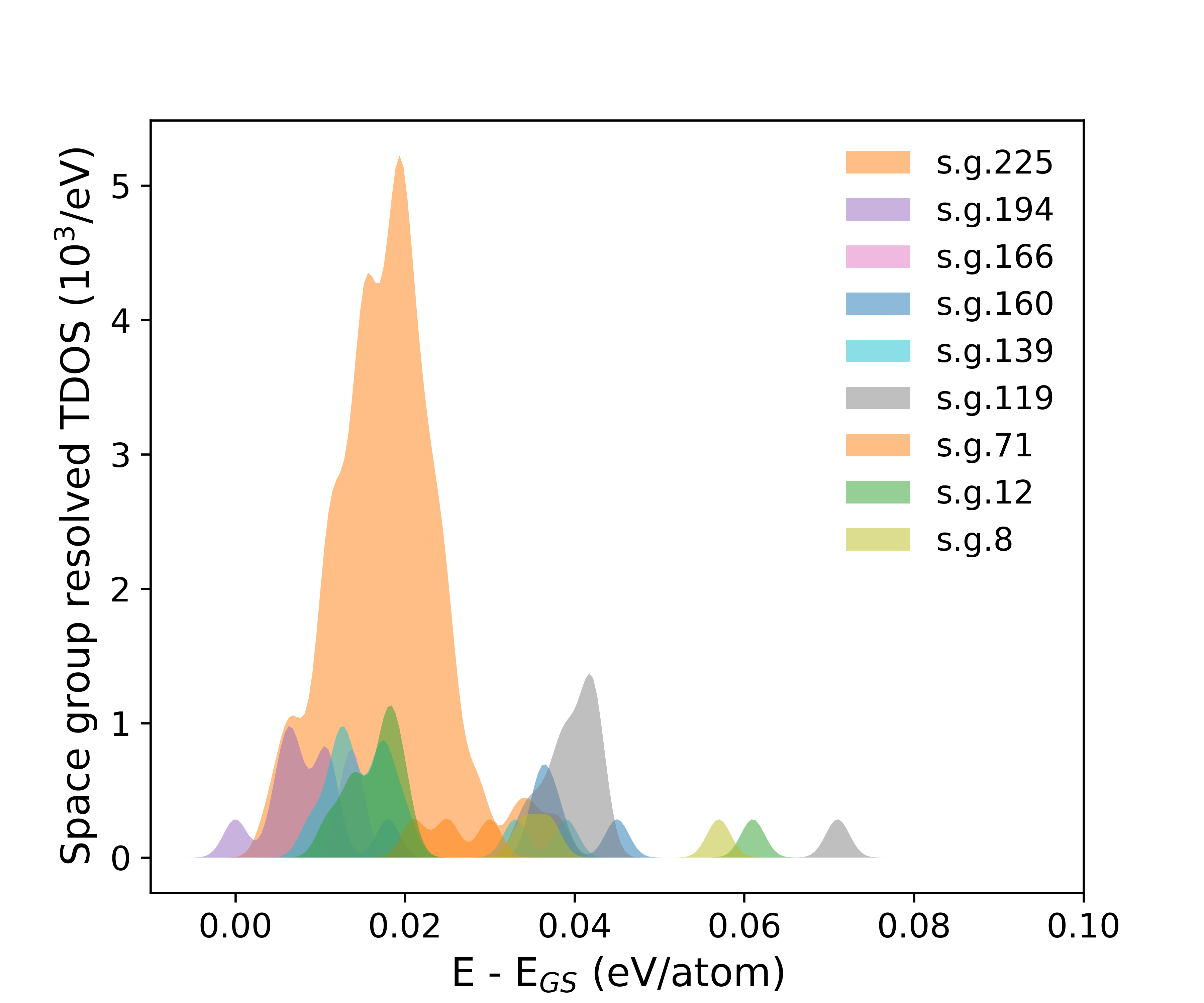}
\caption{\label{MgZnCuO3} The thermodynamic density of states for a hypothetical compound MgZnCuO3. Despite the fact that both Zn and Cu prefer 4-fold coordination in oxides, rocksalt is by far the most abundant structure in random sampling of this compound. This displays the statistical relevance of the rocksalt structure for high-entropy systems. }
\end{figure*}

\begin{figure*}[ht!]
\includegraphics[width=\linewidth]{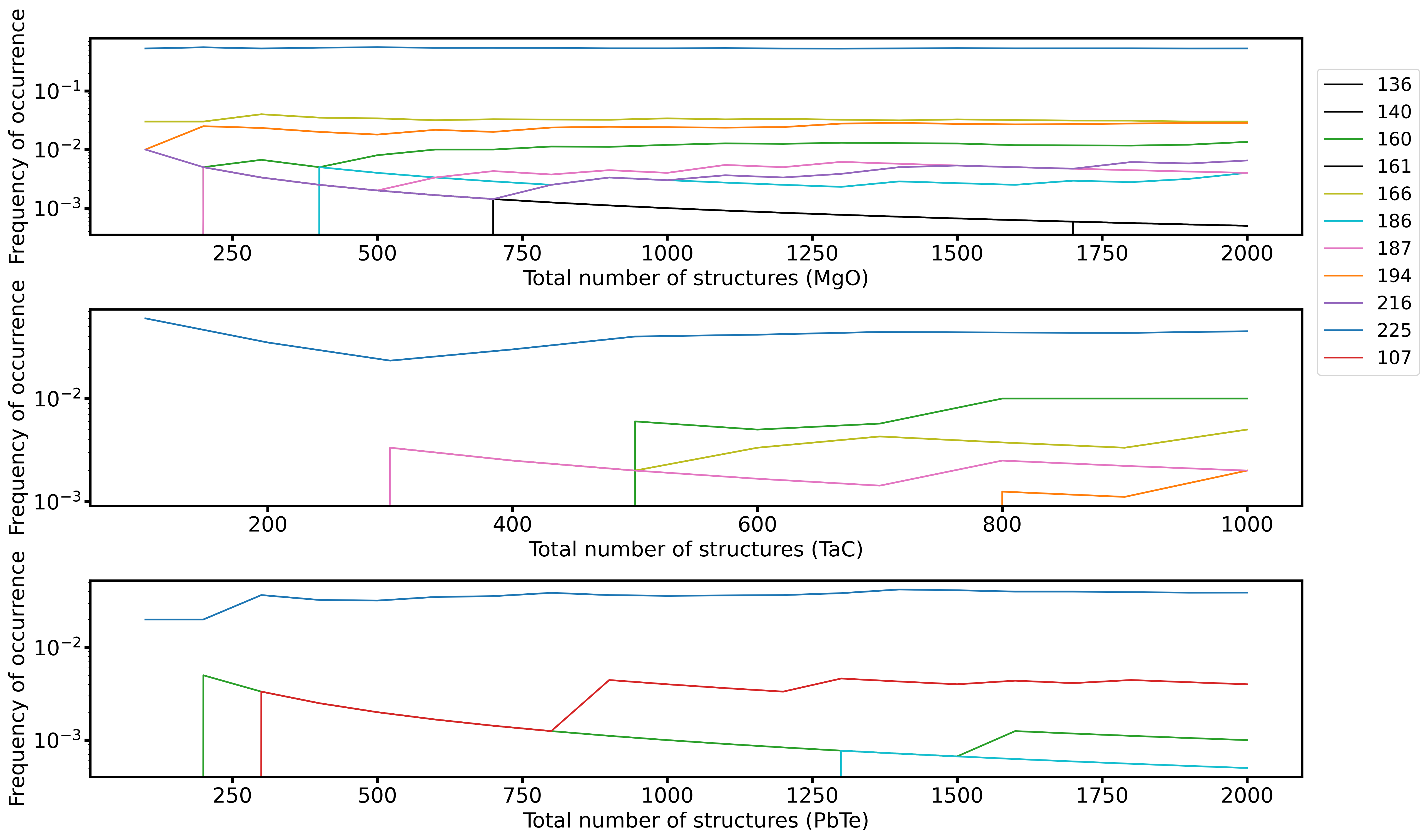}
\caption{\label{poly_sampler_conv} Convergence of the high symmetry structures with respect to total number of structures for random sampling on each of MgO, TaC, and PbTe.}
\end{figure*}

\clearpage

%

\end{document}